# Re-recognition of the ideal gas and real gas

Li Yang [1], WU Jing [2], GUO Zeng-Yuan [1]*

1 Key Laboratory for Thermal Science and Power Engineering of Ministry of Education, Department of Engineering Mechanics, Tsinghua University, Beijing 100084, China

2 Department of Engineering Thermophysics, School of Energy and Power Engineering, Huazhong University of Science and Technology, Wuhan 430074, China

*E-mail: demgzy@tsinghua.edu.cn

**Abstract**: Ideal gas is the most fundamental and simple system in thermodynamics, which has extensive applications in energy research and engineering. By reviewing the physical concept of ideal gas, it is found that the current understanding of ideal gas is still inappropriate and ambiguous, making it challenging to reveal the essential difference between ideal and real gases. Therefore, the macro and microscopic definitions and properties of ideal gas need to be reconceptualized. On the microscopic level, an ideal gas is a hypothetical collection of classical masses in irregular motion, which does not involve quantum mechanics; on the macroscopic level, any real gas cannot strictly obey the equation of state of ideal gas. Moreover, the heat capacity is a constant which should be an endogenous attribute of ideal gas model in order to unify the macroscopic and microscopic definitions of ideal gas. Finally, according to the law of heat capacity, the real gas can be divided into three categories: far-ideal, near-ideal, and quasi-ideal. This makes it easier to perform thermodynamic calculations with the assistance of ideal gas. Among them, the far-ideal gas heat capacity is a function of temperature and pressure, the near-ideal gas heat capacity is a single-valued function of temperature, and the quasi-ideal gas heat capacity is a constant.

# 1　Introduction

Ideal gas is the simplest and most commonly used model in thermodynamics, and it is important to understand its physical meaning accurately in order to understand and analyze the ideal gas equation of state (EOS). The study of the fundamental properties of ideal gas can be traced back to more than 300 years ago, and the most representative ones are Boyle-Mariotte's law[1], Gay-Lussac's law[2] and Charles' law[3]. Based on the summary of the above experimental laws, Clapeyron[4] summarized and derived the original ideal gas EOS (*pV=CT*). Through the research and extension by Dmitri Mendeleev and others[5], the current prevailing ideal gas EOS (*pV=RT*) was obtained, where *R* is the universal gas constant.

The classical ideal gas equation of state encountered a challenge in describing the actual gas properties when Kamerlingh realized in 1908 that for a system consisting of $^4$He atoms. The system can be approximated by the classical ideal gas equation at room temperature, but at low temperatures it cannot be approximated by the ideal gas equation of state at all. After subsequent studies, it was concluded that this is due to the significant quantum effects at low temperatures that amplify the mutual attraction between the particles, thus invalidating the managerial ideal gas equation[6]. For this reason, some researchers divided ideal gas into two categories[7]: classical ideal gas, which belong to the category of managerial physics and can be described by the classical ideal gas equation; and quantum ideal gas, which belong to the category of quantum mechanics and cannot be described by the classical ideal gas equation. So far, the concept of ideal gas seems to have been clearly elaborated and the equation of state of ideal gas has been fully applied in the next[8]. However, a careful reading of the relevant thermodynamic literature and works reveals that there are still misunderstandings in the existing thermodynamic literature regarding the definition and properties of ideal gas.

Turns[9] and Cengel[10] considered from a macroscopic point of view that any gas that follows the the Clausius-Clabellon equation is an ideal gas, which seems to indicate that ideal gas actually exist, while some literature[7, 11] considered ideal gas as hypothetical gas that do not actually exist; Shen[12] and Zeng[13] considered ideal gas as the limit when the gas pressure tends to zero and the specific volume tends to infinity In fact, the ideal gas equation can be satisfied under atmospheric conditions or even at high pressure (as long as the temperature is high); Shen[12], Zeng[13] and Liu[14] considered that the heat capacity of ideal gas is a single-valued function of temperature, however, a careful study of the literature shows that Clausius[15], in his study on the microscopic model of classical ideal gas as early as 1857, already However, a closer look at the literature shows that Clausius[15] pointed out that the heat capacity of a classical ideal gas is a constant as early as 1857 in his study of the microscopic model of the classical ideal gas; most thermodynamic works[12-14, 16] clearly state that the molecules of the classical ideal gas do not occupy space volume themselves, i.e., the molecules have no structure, but the structure of the polyatomic molecules themselves cannot be ignored. It is evident that many thermodynamics textbooks do not clearly state the definition and categories of ideal gas, which leads to misunderstandings about them. Therefore, it is very necessary to review the physical concepts of ideal gas and re-clarify the macroscopic and microscopic definitions of ideal gas.

## 2   Classical ideal gas

Depending on whether quantum effects are significant, ideal gas can be divided into classical ideal gas and quantum ideal gas. When the temperature is low enough or the pressure is high enough, the quantum nature of the gas has become important and the quantum ideal gas undergoes deviations from its classical nature due to quantum mechanical effects. It is important to note that in this study, quantum ideal gas is not discussed.

### 2.1 Microscopic definition

The microscopic definition of the classical ideal gas was determined by Watterston, Klinich and Clausius in the process of developing the theory of motion of heat based on statistical concepts and laws. The unification of the macroscopic and microscopic definitions of the classical ideal gas was not completed until the middle of the 19th century through Clausius, who in 1857[15] first proposed a microscopic model of the classical ideal gas molecule based on the study of the kinetic theory of gas. In order to comply with the gas laws derived from experiments, he considered that the classical ideal gas must satisfy the following conditions.

(1) The space occupied by the molecule itself is infinitely small compared to the space filled by the gas, which means that the molecule can be treated as a mathematical point.

(2) The time taken for one collision between molecules is much smaller than the time interval between two successive collisions.

(3) The molecular force is infinitely small.

Based on the above conditions, the classical ideal gas equation of state can be obtained by derivation from gas kinetic theory. In the system satisfying the above conditions, the relationship between the pressure and the average kinetic energy of the particles is.

$$pV = \frac{1}{3} Nmv_{rms}^2 \tag{1}$$

From Maxwell's distribution:

$$v_{rms}^2 = 4\pi \left(\frac{m}{2\pi kT}\right) \int_0^\infty v^4 e^{-\frac{mv^2}{2kT}} dv \tag{2}$$

After integration is obtained as:

$$v_{rms}^2 = \frac{3kT}{m} \tag{3}$$

Substituting into equation (1), the expression of the classical ideal gas equation of state can be obtained as:

$$p = \frac{NkT}{v} \tag{4}$$

The above derivation must satisfy the assumption of molecular chaos and neglect quantum effects. Meanwhile, by reviewing Clausius' microscopic formulation of the classical ideal gas, it can be found that: (1) since the molecular volume of the actual gas cannot be zero and the intermolecular forces cannot be absent, the classical ideal gas is a hypothetical gas that does not actually exist; (2) since the molecular volume and intermolecular forces of the actual gas are finite values, the actual gas cannot strictly obey the ideal gas (3) the molecules of classical ideal gas have no structure, and they are treated as mathematical points; (4) molecular dynamics theory does not involve quantum mechanics, and classical ideal gas only involve classical theoretical mechanics.

**2.2 Macroscopic definition**

The existing thermodynamic literature usually defines a gas that follows the Clabellon equation obtained based on the experimental law of gas and Avogadro's law as an ideal gas. This definition can easily lead to the misunderstanding that an ideal gas is an actual gas. Precise experiments have shown that the experimental gas law and Arrhenius' law do not fully represent the ideal gas, and that their deviations only decrease as the pressure of the gas decreases. Another macroscopic definition considers an ideal gas as the limiting state when the pressure of the gas tends to zero and the specific volume tends to infinity. This definition is also prone to misunderstandings, as the limit state is unreachable and the relationship between the thermodynamic parameters of the actual gas only approximately satisfies the equation of state of the ideal gas.

In order to further illustrate the difference between the actual gas and the classical ideal gas, we choose the Van der Waals equation of state, which is obtained by considering intermolecular interactions as a correction to the ideal gas equation of state, as an example for analysis.

$$\left(p + \frac{an^2}{V^2}\right)(v - nb) = nRT \tag{5}$$

where $a$ and $b$ are constants for the specific real gas. In the limiting state where the gas pressure tends to zero ($V \to \infty$), the Van der Waals gas approximately satisfies the ideal gas equation of state. It is worth thinking about whether Van der Waals gas is an ideal gas at this point. As an example, the Joule-Thomson coefficient is defined by the equation:

$$\mu = \left(\frac{\partial T}{\partial p}\right)_H = \left(T\left(\frac{\partial v}{\partial T}\right)_P - v\right) \Big/ C_p \tag{6}$$

For ideal gas, the enthalpy is only a single-valued function of temperature, so the temperature is constant before and after throttling, and the joule soup coefficient of ideal gas is zero. For Van der Waals gas, taking the amount of substance $n=1$mol, substituting equation (7) into equation (8) and combining with the thermodynamic cycle relation equation, the Joule-Thomson coefficient is expressed as:

$$\mu = \frac{1}{C_p}\left(\frac{-RTbv^3 + 2av(v-b)^2}{nRTv^3 - 2a(v-b)^2}\right) \tag{7}$$

Any equation of state is only compatible with reality within a certain range of parameters. The van der Waals gas equation with deterministic coefficients, in the large volume limit, gradually deviates from the actual gas and tends to the ideal state.

When the volume of van der Waals gas tends to the limit, the Joule-Thomson coefficient takes the limit as:

$$\lim_{v \to \infty} \mu \approx \frac{1}{C_p}\left(-b + \frac{2a}{RT}\right) \quad (8)$$

For the van der Waals gas in the limit state, the van der Waals equation of state degenerates to the ideal gas equation of state when and only when the van der Waals coefficients a and b are both zero. In other words, for van der Waals gas with non-zero van der Waals coefficients, when the volume tends to infinity, the van der Waals gas only approximately satisfies the ideal gas equation of state, but it is still not an ideal gas. In short, ideal gas do not exist, and no real gas can reach the limiting state or strictly follow the equation of state of an ideal gas.

**3 Heat capacity of classical ideal gas**

Within the scope of classical statistical mechanics, macroscopic-based and microscopic-based analyses are prone to some contradictions in the laws governing the variation of heat capacity of ideal gas. From the macroscopic point of view, there should be two possibilities to characterize the heat capacity obtained from the differentiation of internal energy with respect to temperature, i.e., the heat capacity can be either a function of temperature or a constant. From the microscopic point of view, Clausius, who was the first to propose the microscopic model of the classical ideal gas, has clearly pointed out in his own publication that the heat capacity of the classical ideal gas is constant, which can be confirmed by the energy homogeneity theorem. For the ideal gas obeying the classical statistical theory, the molecule can be regarded as a mass, and the total energy of the ideal gas in each microstate is the sum of the affine kinetic energy of each molecule. The molecules of the ideal gas have only advective motion and their energies are：

$$\varepsilon = \frac{1}{2m}\left(p_x^2 + p_y^2 + p_z^2\right) \quad (9)$$

According to the energy equalization principle, the constant heat capacity of an ideal gas is：

$$C_V = \left(\frac{\partial U}{\partial T}\right)_V = \frac{3}{2}Nk \quad (10)$$

The above analysis shows that the heat capacity of an ideal gas is a constant in the classical mechanics, and the heat capacity of an ideal gas based on the classical statistical theory is obtained by further idealization of the actual gas. Since the above classical theoretical values do not correspond to the experimental measurements of the heat capacity of the actual gas. This, together with the common misinterpretation of the actual gas, which is not too low in temperature and not too high in pressure, as an ideal gas, makes people think that the classical theory does not fully describe the thermal property law of the ideal gas. In order to be consistent with experimental observations, quantum theory is introduced to describe the classical ideal gas in the microscopic state. At the same time, the misunderstanding of the macroscopic definition of the classical ideal gas has led to the misconception that the heat capacity of the dilute real gas is also the heat capacity of the classical ideal gas, and thus to the further misconception that the heat capacity of the classical ideal gas is a single-valued function of temperature.

The analysis is now carried out in another way: since the microscopic definition of the classical ideal gas is unified with the macroscopic definition, the heat capacity characteristics derived from the two analyses should not be contradictory. In the context of classical statistical theory, the heat capacity of a classical ideal gas must also be constant. Therefore, in order to unify the macroscopic definition of the classical ideal gas with the microscopic statistical theory, the heat capacity being constant should also be an endowment property of the macroscopic model of the classical ideal gas. In other words, in addition to strictly following the ideal gas equation of state, the classical ideal gas requires the heat capacity to be constant.

**4 Classification of ideal gas**

The lack of clarity in the definition of ideal gas in thermodynamics can lead to misunderstandings about the properties of ideal gas, and this lack of clarity often comes from the lack of clarity in the classification of ideal gas. In the case of heat capacity, for example, since most works on thermodynamics refer to it generically as an ideal gas, heat capacity is sometimes treated as a constant and sometimes as a function of temperature. In both cases, the process equations as well as the thermodynamic functions are formulated differently, sometimes causing confusion and even errors.

To this end, as shown in Fig. 1, based on the above analysis and combining classical and quantum statistical theories, ideal gas can be defined as classical and semiclassical ideal gas, respectively. Those that can be described by the ideal gas equation of state are classical and semiclassical ideal gas; those that cannot be described by the ideal gas equation of state are quantum ideal gas. In the case of classical ideal gas, in addition to the strict compliance with the equation of state of ideal gas, the macroscopic heat capacity is also required to be constant; in the microscopic case, it is a classical collection of masses without interaction forces, and its properties are explained by classical statistical theory. In the case of semiclassical ideal gas, the macroscopic one only requires strict compliance with the ideal gas equation of state, and the heat capacity is a single-valued function of temperature; the microscopic one consists of particles without interaction, and its properties are explained by quantum statistical theory. The above classification does not increase the length of the textbook, but the previously mentioned confusion will naturally be clarified.

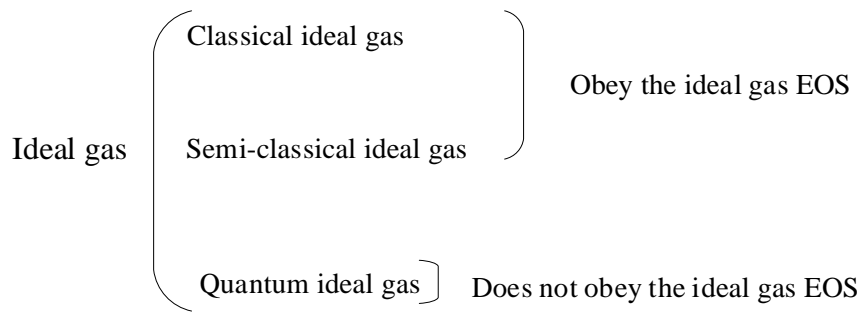

Fig. 1 Classification of ideal gas

## 5 Actual gas

Based on the knowledge of the classical ideal gas in the previous section, the microscopic and macroscopic definitions of the actual gas can be easily clarified. In terms of the microscopic definition: the real gas has a non-negligible molecular structure, molecular volume and molecular interactions; in terms of the macroscopic definition: the real gas does not satisfy the ideal gas equation of state (approximately in the limit) and its heat capacity is a function of temperature or pressure. It is because molecules have volume, structure and intermolecular forces that the heat capacity of a real gas must not be constant.

5.1 Characteristics of actual gas

Heat capacity as a function of temperature or pressure is one of the characteristics that distinguish real gas from ideal gas. The vague understanding of the macroscopic definition of classical ideal gas makes people usually mistake the heat capacity of dilute real gas for the heat capacity of classical ideal gas, thus further mistaking the heat capacity of classical ideal gas as a single-valued function of temperature. In addition, since the existing literature usually uses both classical and quantum statistical methods to calculate the heat capacity of real gas, it makes people confuse the understanding of the thermal properties of classical ideal gas at the macroscopic level as well. By reconceptualizing the microscopic and macroscopic definitions of the classical ideal gas, it is clear that the classical ideal gas does not exist and it does not involve the category of quantum statistics.

For single-atom real gas, the electrons inside the atoms are frozen in the ground state at room temperature and do not contribute to the heat capacity, which is the same as that obtained by quantum and classical statistics. For polyatomic real gas, the structure of the molecules themselves cannot be neglected and the heat capacity needs to be analyzed using quantum statistical theory. In the case of diatomic real gas, for example, the advection, rotation and vibration terms of the molecules all contribute to the heat capacity of the gas. The contribution of advection to the heat capacity can be obtained directly from the advection partition function as $3Nk/2$, which is consistent with the results obtained from the classical statistical energy parity theorem. The relative vibration of two atoms in a hydrogen molecule can be approximated as a linear resonant oscillator, whose vibrational partition function is shown as follows.

$$Z_1^V = \frac{e^{-\frac{\beta \hbar \omega}{2}}}{1 - e^{-\beta \hbar \omega}} \tag{11}$$

Thus the contribution of vibration to the internal energy is:

$$U^V = -N \frac{\partial}{\partial \beta} \ln Z_1^V = \frac{N \hbar \omega}{2} + \frac{N \hbar \omega}{e^{\beta \hbar \omega} - 1} \tag{12}$$

The first term in the above equation is the zero-point energy of $N$ oscillators, independent of temperature; the second term is the thermal excitation energy of $N$ oscillators at a temperature of $T$. After introducing the vibration characteristic temperature $k\theta_v = \hbar\omega$, the contribution of vibration to the constant heat capacity is:

$$C_V^V = Nk\left(\frac{\theta_v}{T}\right)^2 \frac{e^{\theta_v/T}}{\left(e^{\theta_v/T}-1\right)^2} \tag{13}$$

For the hydrogen molecule, the vibrational characteristic temperature is 103 K. Due to the energy level discrete, the oscillator must acquire energy ℏω to be able to jump to the excited state. Therefore, when the temperature is low ($T \ll \theta v$), almost all the oscillators are frozen in the ground state and the contribution of the vibrational degrees of freedom to the heat capacity is close to zero; when the temperature is extremely high ($T \gg \theta v$), the vibrational contribution to the fixed-capacity heat capacity is $Nk$.

When considering the rotation of a diatomic molecule, the rotation characteristic temperature $k\theta_r = \hbar^2/2l$ is introduced. with the help of the rotation energy level and the rotation simplicity, the rotation partition function is:

$$Z_1^r = \sum_{l=0}^{\infty}(2l+1)e^{-\frac{\theta_r}{T}l(l+1)} \tag{14}$$

In the above equation, $l$ is the quantum number of rotation. In the room temperature range, the rotation energy spacing is much less than $kT$, i.e., $\theta_r \ll T$. Therefore, $(l^2+1)\theta_r/T$ can be approximated as a quasi-continuous variable and the contribution of rotation to the heat capacity can be found as:

$$C_V^r = Nk \tag{15}$$

In summary, the diatomic real gas heat capacity obtained based on the quantum statistical method is approximated as:

$$C_V = \frac{3}{2}Nk + Nk + Nk\left(\frac{\theta_v}{T}\right)\frac{e^{\theta_v/T}}{(e^{\theta_v/T}-1)^2} \tag{16}$$

Under the resonant oscillator approximation, the heat capacity curve of the hydrogen molecule is approximated as shown in Fig. 2 by substituting the rotation and vibration characteristic temperature information of the hydrogen molecule. The rotational and vibrational degrees of freedom excited states of the molecule are different at different temperatures, which is the effect caused by quantum mechanical effects. At low temperatures, the system reflects only the translational degrees of freedom, i.e., the molecules do not rotate or vibrate, and the diatomic gas heat capacity is $3Nk/2$. When the temperature rises to a certain level, the rotational state of the molecules is fully excited, and the hydrogen gas heat capacity is 5*Nk*/2. At higher temperatures, the vibrational degrees of freedom are excited, and the hydrogen gas heat capacity when fully excited is 7*Nk*/2. It can be seen that when the system of actual gas is at a very high temperature, all the energy of its rotation and vibration has been excited and the heat capacity is tending to be a constant independent of temperature and pressure.

Classical statistics are the limit of quantum statistics and, accordingly, classical ideal gas are the limit states of real gas. For a system of molecules of a real gas, the condition that it satisfies classical statistics can be obtained by solving for its thermal wavelengths. Among them, the thermal de Broglie wavelength is:

$$\lambda = h \bigg/ \int_{-\infty}^{\infty} \exp\left(-\frac{p^2}{2m}\right) dp = h \bigg/ \sqrt{2\pi kTm} \qquad (17)$$

When the molecular distance is much larger than the average thermal wavelength of the molecules:

$$(V/N)^{1/3} \gg 1 \qquad (18)$$

or

$$\frac{V}{N}\left(\frac{2\pi mkT}{h^2}\right)^{3/2} \gg 1 \qquad (19)$$

From the above equation, it is clear that if the gas is more dilute, the higher the temperature and the larger the mass of the molecules, the more easily the classical conditions are satisfied for a defined gas system. Obviously, in the limit state, the thermodynamic parameter relations tend to the ideal gas equation of state, and the heat capacity tends to be a constant independent of temperature and pressure. This indicates that the diatomic real gas, like the monatomic real gas, tends to have the properties of the classical ideal gas in the limiting parameter conditions, and that the thermodynamic parameter relations of the real gas can be approximated by the ideal gas equation of state when the pressure is not too high and the temperature is not too low.

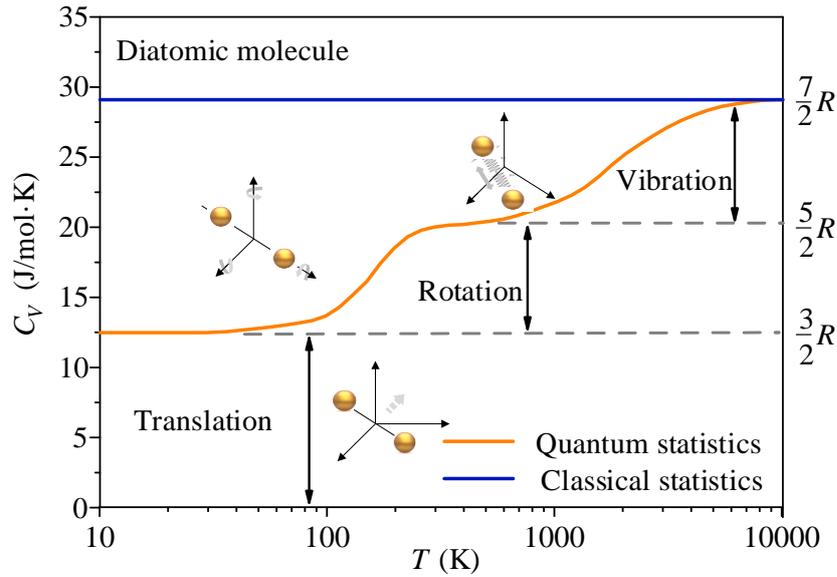

Fig. 2 Variation of heat capacity of hydrogen gas under different statistical methods

5.2 Classification of actual gas

In the field of hypersonic velocity, in order to facilitate the study of compressible gas flow problems, the actual gas are usually classified into calorimetric and thermally complete gas based on whether the heat capacity is a function of temperature, and many formulas and calculations describing the laws of motion of compressible gas have been derived accordingly. Obviously, the classification of real gas according to their properties has important application value, however, in the field of thermodynamics, the classification according to the properties of heat capacity has not yet been studied in the literature.

To make it more convenient to calculate the thermodynamic properties of real gas with the help of classical ideal gas, we can classify the real gas according to the characteristics of its states and heat capacity. As shown in Table 1, the real gas can be classified into three categories: far-ideal gas, near-ideal gas, and quasi-ideal gas, which are the differences between different real gases, not the differences between real and ideal gas. Unlike the conventional thermodynamics, where the compression factor is used to express the deviation of actual gas from ideal gas, the heat capacity is used to classify actual gas in Table 1. According to the characteristics of heat capacity, real gas can be further classified as quasi-ideal gas with approximate constant heat capacity, near-ideal gas with approximate heat capacity as a single-value function of temperature, and far-ideal gas with heat capacity as a function of temperature and pressure. Any real gas can be classified into the above three categories of gas according to their thermodynamic parameter ranges, but the corresponding parameter ranges are different for different gas.

Table. 1 Classification of actual gas

|  | **quasi-ideal gas** | **near-ideal gas** | **far-ideal gas** |
| --- | --- | --- | --- |
| Ideal gas EOS | Approximate satisfaction | Approximate satisfaction | Not satisfaction |
| Heat capacity | $C$ = constant | $C = C(T)$ | $C = C(T, p)$ |
| Internal energy | $U = U(T)$ | $U = U(T)$ | $U = U(T, p)$ |

For monatomic gases, in a given region (10K < $T$ < 1000K, 10kPa < $p$ < 1200kPa), the actual gas is classified as far-ideal, near-ideal and quasi-ideal with the help of the degree of deviation from the ideal gas and the magnitude of the relative error between the heat capacity and the heat capacity of the actual gas at zero pressure, as shown in Eq. (20).

$$\begin{cases} 1\% \leq \dfrac{C_p - C_{p=0}}{C_{p=0}} & \text{Far-ideal gas} \\ 0.01\% \leq \dfrac{C_p - C_{p=0}}{C_{p=0}} < 1\% & \text{Near-ideal gas} \\ \dfrac{C_p - C_{p=0}}{C_{p=0}} < 0.01\% & \text{Quasi-ideal gas} \end{cases} \quad (20)$$

Fig. 3 depicts the range of regions for the three types of helium for different thermodynamic parameters. As shown in Fig. 3(a), the effect of pressure on the heat capacity decreases as the temperature increases, and when the temperature increases to a certain value, the heat capacity tends to be constant. As shown in Fig. 3(b), for a given pressure condition ($p$ = 1.5 kPa), when 2.17 K < $T$ < 3.84 K, the heat capacity is affected by both temperature and pressure, and the helium in this region is called far ideal gas; when 3.84 K < $T$ < 31.18 K, the pressure has a negligible effect on the heat capacity, and the heat capacity is mainly affected by temperature, and the helium in this region is called near ideal gas When 31.18K < $T$, the heat capacity can be approximated as a constant and the helium in this region can be called quasi-ideal.

For polyatomic real gas, the molecular structure is not negligible and the degrees of freedom of the molecule is potentially quantized at different temperatures. Therefore, in this region (50 K < $T$ < 1000 K, 10 kPa < $p$ < 1200 kPa), the polyatomic gas can be classified into far-ideal and near-ideal gas states, as shown in Eq. (21).

$$\begin{cases} 1\% \leq \dfrac{C_p - C_{p=0}}{C_{p=0}} & \text{Far-ideal gas} \\ \dfrac{C_p - C_{p=0}}{C_{p=0}} < 1\% & \text{Near-ideal gas} \end{cases} \tag{21}$$

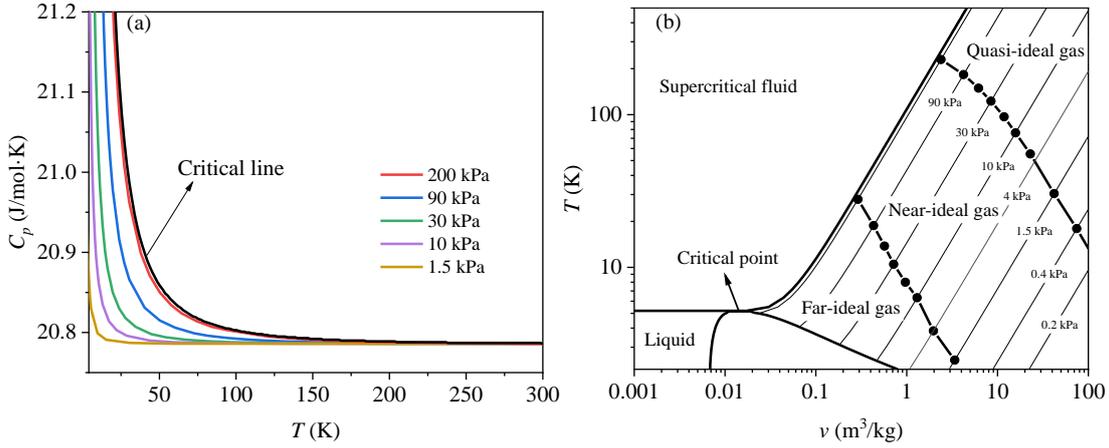

Fig. 3 Regional ranges of the three types of helium

Fig. 4 depicts the range of regions for the three types of hydrogen gases for different thermodynamic parameters. As shown in Fig. 4(a), the heat capacity of the polyatomic gas decreases gradually by pressure as the temperature continues to increase. As shown in Fig. 4(b), for a given pressure working condition ($p = 100$ kPa), when 20.45 K $< T <$ 63.96 K, the hydrogen in this region is called far ideal gas; when $T >$ 63.96 K, the hydrogen in this region is called near ideal gas.

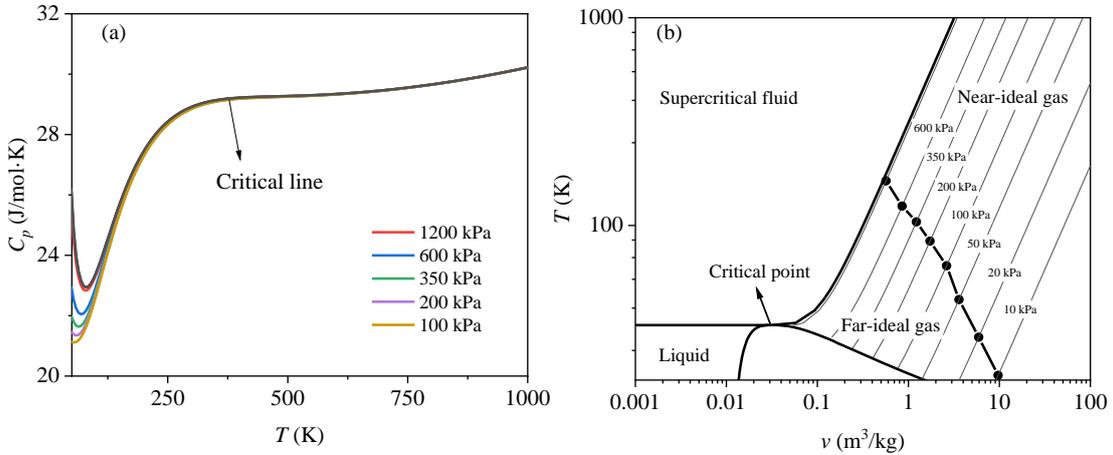

Fig. 4 Regional ranges of the three types of hydrogen

**6 Conclusion**

(1) "Ideal" is the simplest concept that ignores various practical factors. An ideal gas is a hypothetical gas that does not actually exist. The actual gas parameters at low pressure or small volume only approximately satisfy the equation of state of the ideal gas, and the state parameters of any actual gas cannot strictly follow the equation of state of the ideal gas.

(2) Ideal gases can be classified as quantum ideal gases, semiclassical ideal gases and classical ideal gases. From the microscopic point of view, the classical ideal gas has no volume, no structure, no interaction between molecules, and only involves classical statistics. From the macroscopic point of view, the classical ideal gas not only follows strictly the ideal gas equation of state, but also its heat capacity should be constant.

(3) Based on the characteristics of the heat capacity of the real gas, the real gas can be divided into three categories. The heat capacity of quasi-ideal gases is approximately constant, the heat capacity of near-ideal gases is approximately a single-valued function of temperature, and the heat capacity of far-ideal gases is a function of temperature and pressure. In addition, the ideal gas equation can approximately describe the relationship between the state parameters of quasi-ideal and near-ideal gases. Any real gas can be classified into the above three categories of gases according to their thermodynamic parameter range.